\documentclass[aps,prb,twocolumn]{revtex4}

\newcommand{\kk}{{\bf k}}
\usepackage[]{graphicx}
\usepackage{amsbsy,amsmath,amssymb,amsthm,amsfonts}
\usepackage{times}
\usepackage{bm}
\usepackage{units}

\newcommand{\comment}[1]{}

\begin{document}

\title{Review on carrier multiplication in graphene}

\author{Ermin Malic$^1$}
\email[]{ermin.malic@chalmers.se}
\author{Torben Winzer$^2$}
\author{Florian Wendler$^2$}
\author{Andreas Knorr$^2$} 

\affiliation{$^1$ Department of Physics, Chalmers University of Technology, 41296 Gothenburg, Sweden}
\affiliation{$^2$ Department for Theoretical Physics, Technical University Berlin, 10623 Berlin, Germany}

\begin{abstract}
The remarkable gapless and linear band structure of graphene opens up new carrier relaxation channels bridging the valence and the conduction band. These Auger scattering processes change the number of charge carriers and can give rise to a significant multiplication of optically excited carriers in graphene. This is an ultrafast many-particle phenomenon that is of great interest both for fundamental many-particle physics as well as technological applications. Here, we review the research on carrier multiplication in graphene and Landau-quantized graphene including theoretical modelling and experimental demonstration.
\end{abstract}

\maketitle

\begin{figure}[t!]%
\centering
\includegraphics*[width=\columnwidth]{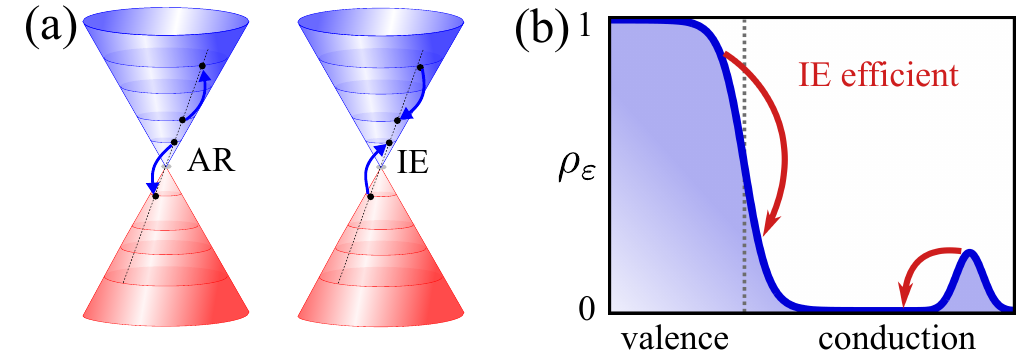}
\caption{ Schematic illustration of Auger scattering processes including 
Auger recombination (AR) and the inverse process of impact excitation (IE). The latter can give rise to a carrier multiplication. Figure adapted from Ref. \cite{malic13} and Ref. \cite{winzer12b}. }
\label{fig1_sketch}
\end{figure}

The most fascinating ultrafast phenomenon characterizing the carrier dynamics in graphene is the appearance of carrier multiplication (CM), i.e. generation of multiple electron-hole pairs through internal scattering \cite{nozik02}. 
The underlying physical mechanism of this phenomenon are Auger processes, which are specific Coulomb interband scattering events, where one carrier (electron or hole) bridges the valence and the conduction band, while the other involved carrier remains in the same band, cf. Fig. \ref{fig1_sketch}(a). As a result, Auger scattering changes the overall charge carrier density consisting of electrons in the conduction and holes in the valence band. We distinguish Auger recombination (AR) and the inverse process of impact excitation (IE) \cite{malic13}. The latter inreases, while the first reduces the number of carriers. The efficiency of the two Auger processes depends on the excitation regime and the resulting Pauli blocking. 
After a weak optical excitation, the probability of the interband process close to the Dirac point is higher for IE, since here an electron from the almost full valence band scatters into the weakly populated conduction band, cf.  Fig. \ref{fig1_sketch}(b). The inverse process of AR is strongly suppressed by Pauli blocking. As a result, carrier multiplication can take place \cite{winzer10,winzer12b}.

The theoretical prediction of carrier multiplication for
semiconductor quantum dots in 2002 \cite{nozik02} and its experimental
verification two years later \cite{schaller04} has attracted much
attention \cite{schaller05,ellingson05,gur05,scholes06,schaller06,nair11}. Since then, the concept has been extended to other nanomaterials including carbon nanotubes and graphene \cite{gabor09,baer10,wang10,winzer10,gabor13,kanemitsu13,brida13,wendler14,ploetzing14}.
This ultrafast phenomenon has the potential
to increase the power conversion efficiency of single-junction solar
cells \cite{landsberg93} above the Shockley-Queisser limit \cite{shockley61}.
Moreover, carrier multiplication enables a fast detection of photons
with a high responsivity. In graphene, the occurrence of carrier multiplication
has been recently theoretically predicted \cite{winzer10,winzer12b,basko13,kadi15}
and experimentally confirmed \cite{brida13,ploetzing14,gierz15,hofmann15}. 
Furthermore, Landau-quantized graphene exhibiting an externally tunable bandgap has been suggested 
to exploit this effect for designing graphene-based photovoltaic devices. In spite of the non-equidistant Landau quantization, Auger scattering has been theoretically and experimentally demonstrated to be the crucial relaxation channel \cite{winnerl15} resulting in a carrier multiplication \cite{wendler14}.

To microscopically access the ultrafast carrier dynamics in graphene including the phenomenon of carrier multiplication, we have developed a theoretical approach based on graphene Bloch equations within the second-order Born Markov approximation \cite{knorr96,malic13,malic11b,malic16,wendler15}. This system of coupled equations of motion for carrier occupation $\rho^{\lambda}_{\bf k}(t)$, microscopic polarization $p_{\bf k}(t)$, and phonon number $n_{\bf q}^j$
include contributions from carrier-light, carrier-carrier, and carrier-phonon interaction on a consistent microscopic footing. Here, the electronic and phonon momentum are denoted by $\bf k$ and $\bf q$, furthermore, $\lambda=c,v$ stands for the conduction (c) or valence (v) band, and $j$ describes the considered optical or acoustic phonon modes.
Solving the graphene Bloch equations, we have access to the temporal evolution of the carrier density $\mathcal{N}(t)$ at different excitation conditions allowing us to address the question whether and under which conditions carrier multiplicaiton takes place in graphene.  The many-particle-induced carrier generation can be quantified by a time-dependent carrier multiplication factor $\rm{CM}(t)$
\begin{equation}
 \rm{CM}(t)=\frac{\mathcal{N}(t)}{\mathcal{N}_{opt}(t)},\label{eq_CMdef}
\end{equation}
 which is defined as the ratio of the generated carrier density $\mathcal{N}(t)$ and the purely optically excited density $\mathcal{N}_{opt}(t)$.

\begin{figure}[t!]%
\centering
\includegraphics*[width=0.85\columnwidth]{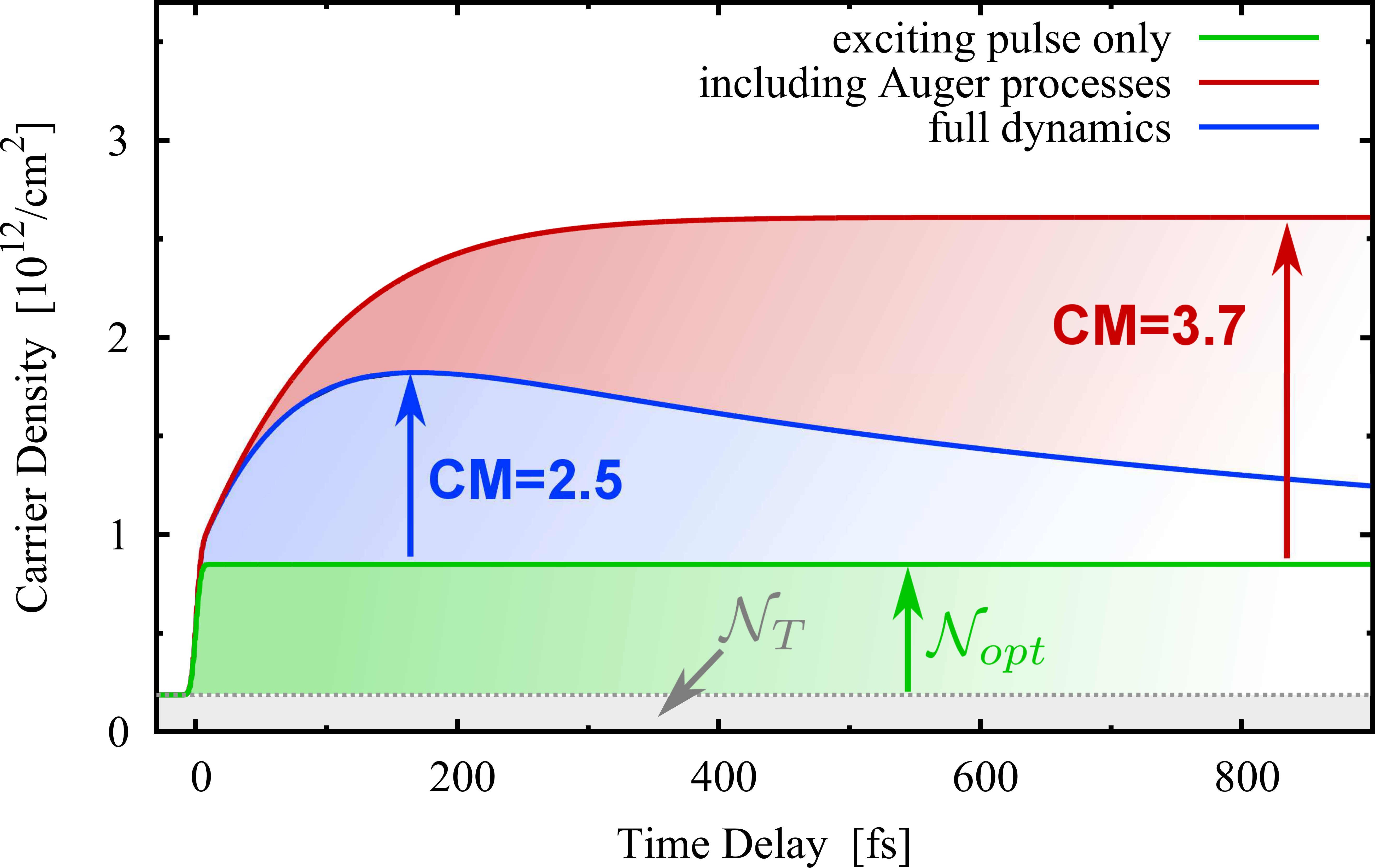}
\caption{Temporal evolution of the carrier density accounting for all relaxation channels (blue line), only carrier-light interaction (green line), and only  Coulomb scattering (red). Auger scattering leads to a multiplication of the initially excited carrier density $\mathcal{N}_{\rm{opt}}$ by a factor of $3.7$. Phonon-induced recombination counteracts the multiple charge carrier generation and reduces the CM factor to $2.5$. Figure taken from Ref. \cite{torben_phd}}.
\label{fig2_cm}
\end{figure}

\section{Carrier multiplication in undoped graphene}
Here, we investigate the ultrafast carrier dynamics in undoped graphene. We find a significant increase of the carrier density $\mathcal{N}(t)$ after optical excitation, cf. Fig. \ref{fig2_cm}. To understand the underlying microscopic mechanism, we subsequently switch on different scattering channels. 
Accounting only for carrier-light interaction, the applied ultrashort optical pulse lifts electrons from the valence into the conduction band resulting in an optically induced carrier density $\mathcal{N}_{\rm{opt}}(t)$,  which remains constant after the pulse is switched off (green line). Taking into account Coulomb-induced scattering channels including Auger processes, we find a further increase of the carrier density even after the pulse has been switched off (red line) - a carrier multiplication takes place.  In the case of a purely Coulomb-driven dynamics, we obtain a multiplication factor of $\rm{CM}=3.7$. However, the 
dynamics is still uncomplete until we have included phonon-induced relaxation channels. These are important since 
they directly compete with Auger processes: an excited electron can scatter to an energetically lower state by emitting a phonon and thus reducing the efficiency of Auger scattering.  Furthermore, phonon-induced interband processes can directly change the carrier density. Since non-radiative recombination involving the emission of phonons is efficient, it reduces the carrier density counteracting the effect of CM. In spite of these effects, taking the full dynamics into account we still observe a carrier multiplication with a maximum value of approximately $\rm{CM}=2.5$ that persists on a picosecond timescale, cf. the blue line in Fig. \ref{fig2_cm} \cite{winzer10,winzer12b,winzer14}. 

\begin{figure}[t!]%
\centering
\includegraphics*[width=0.75\columnwidth]{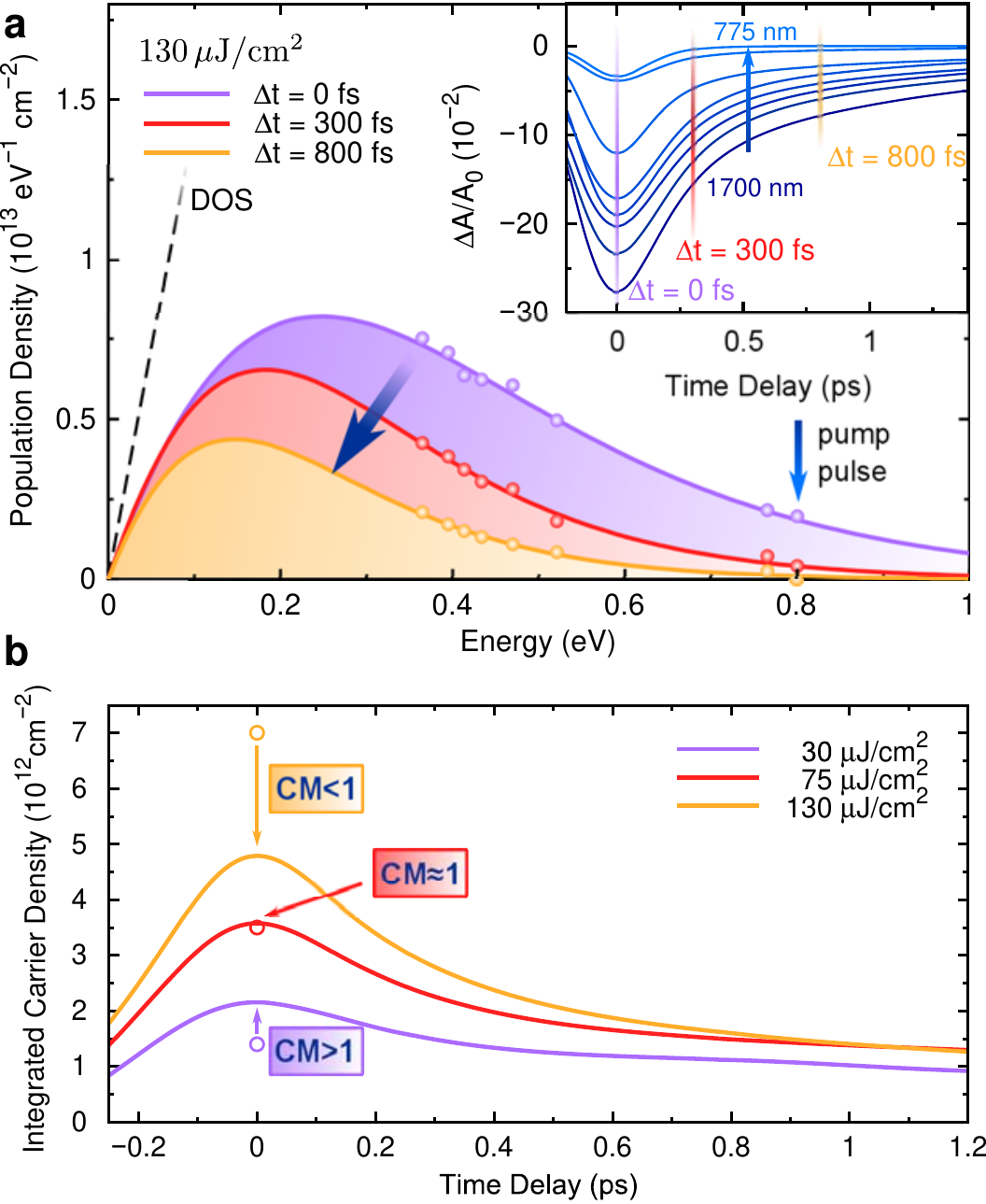}
\caption{(a) Thermalized carrier distributions at three different times,
extracted from pump-probe measurements of the differential absorption $\Delta A/{A_0}$
(see inset) recorded at eight different probe wavelength for an
excitation fluence of $\unit[130]{\mu J/cm^2}$.  (b) Temporal
evolution of the integrated carrier density after optical excitation at
three different excitation fluences. The circles denote the estimated optically excited
carrier density extracted from the applied pump fluence and the
measured constant absorption of graphene on sapphire.  Figure taken from Ref. \cite{ploetzing14}.}
\label{fig3_exp}
\end{figure}
The theoretically predicted carrier multiplication in graphene could be recently also demonstrated in high-resolution multi-color pump-probe \cite{ploetzing14,brida13} and time-and angle-resolved
photoemission (ARPES) measurements \cite{gierz15,hofmann15}.
Direct experimental access to the temporal evolution of the carrier
density in graphene is a challenging task. As absorption at optical frequencies is dominated by
interband processes, pump-probe experiments can be used to 
directly monitor carrier occupation probabilities of the
optically coupled states as well as the Coulomb- and phonon-induced
carrier escape from these states. Ploetzing an co-workers \cite{ploetzing14} performed a series of pump-probe measurements applying different optical probe energies ranging from 0.73 to 1.6 eV, while the pump pulse was fixed at 1.6 eV. Assuming quasi-instantaneous thermalization of the excited carriers through ultrafast carrier-carrier scattering, the measured occupation probabilities at different distinct energies could be exploited to reconstruct the time-dependent
carrier distribution in the relevant range in the momentum space, cf. Fig. \ref{fig3_exp}(a). To be able to draw qualitative conclusions on the appearing carrier multiplication, the optically excited carrier density $\mathcal{N}_{opt}$ is estimated based on the pump fluence and the dark absorption of graphene on sapphire. Depending on the pump fluence, $\mathcal{N}_{opt}$ lies below or above the intergrated carrier density $\mathcal{N}$, cf. Fig. \ref{fig3_exp}(b). This is a clear experimental prove of the appearance of CM in graphene. 

\begin{figure}[t!]%
\centering
\includegraphics*[width=0.75\columnwidth]{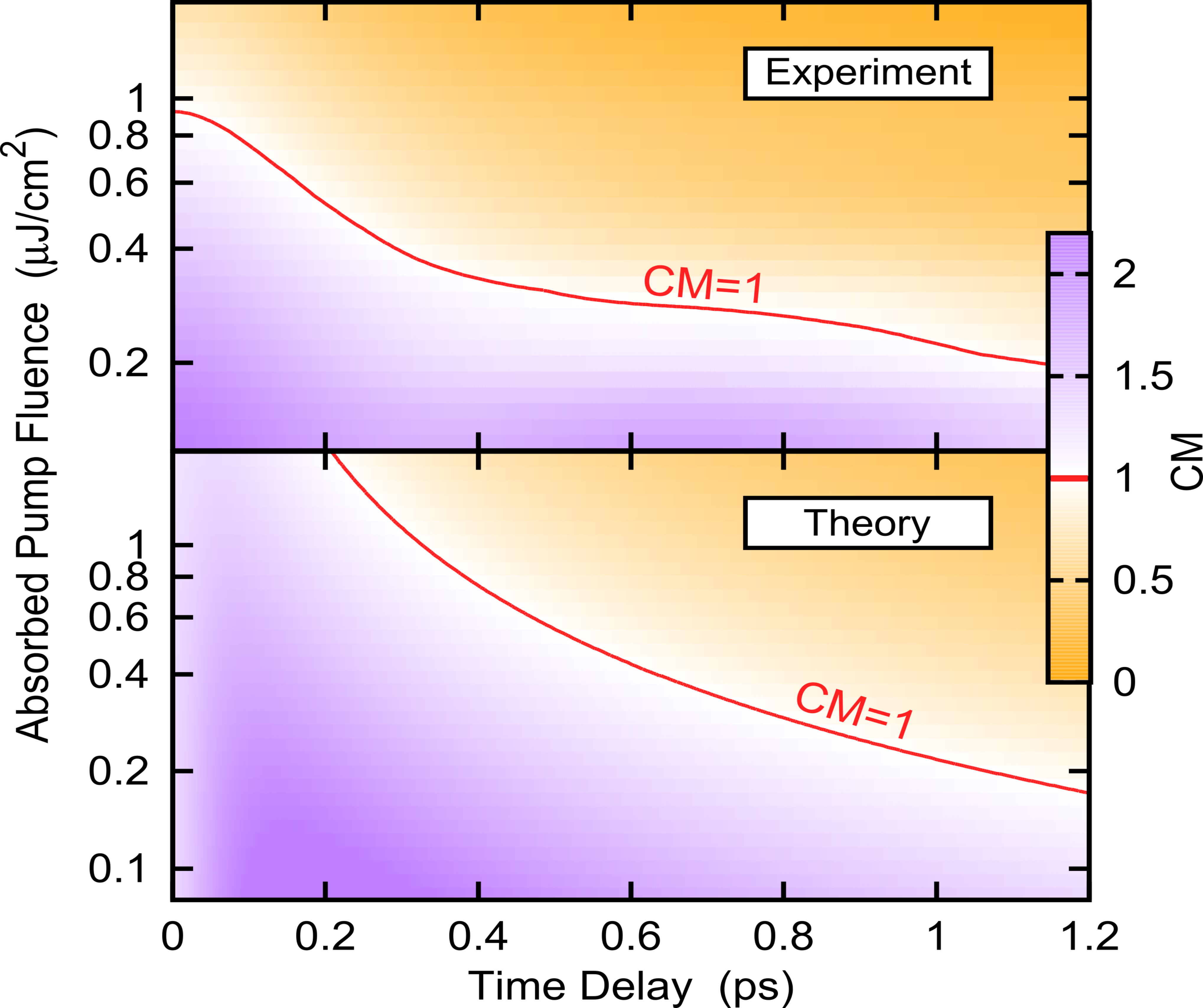}
\caption{Direct comparison of theoretically predicted and experimentally measured carrier multiplication. Purple colors mark the region characterized by a CM
factor larger than 1, where the carrier density exceeds the density of optically injected carriers. Both in experiment  and theory maximum carrier multiplication occurs at low fluences on a timescale of a few picoseconds. Figure taken from Ref. \cite{ploetzing14}. }
\label{fig4_cm_exp}
\end{figure}

\begin{figure}[b!]%
\centering
\includegraphics*[width=0.75\columnwidth]{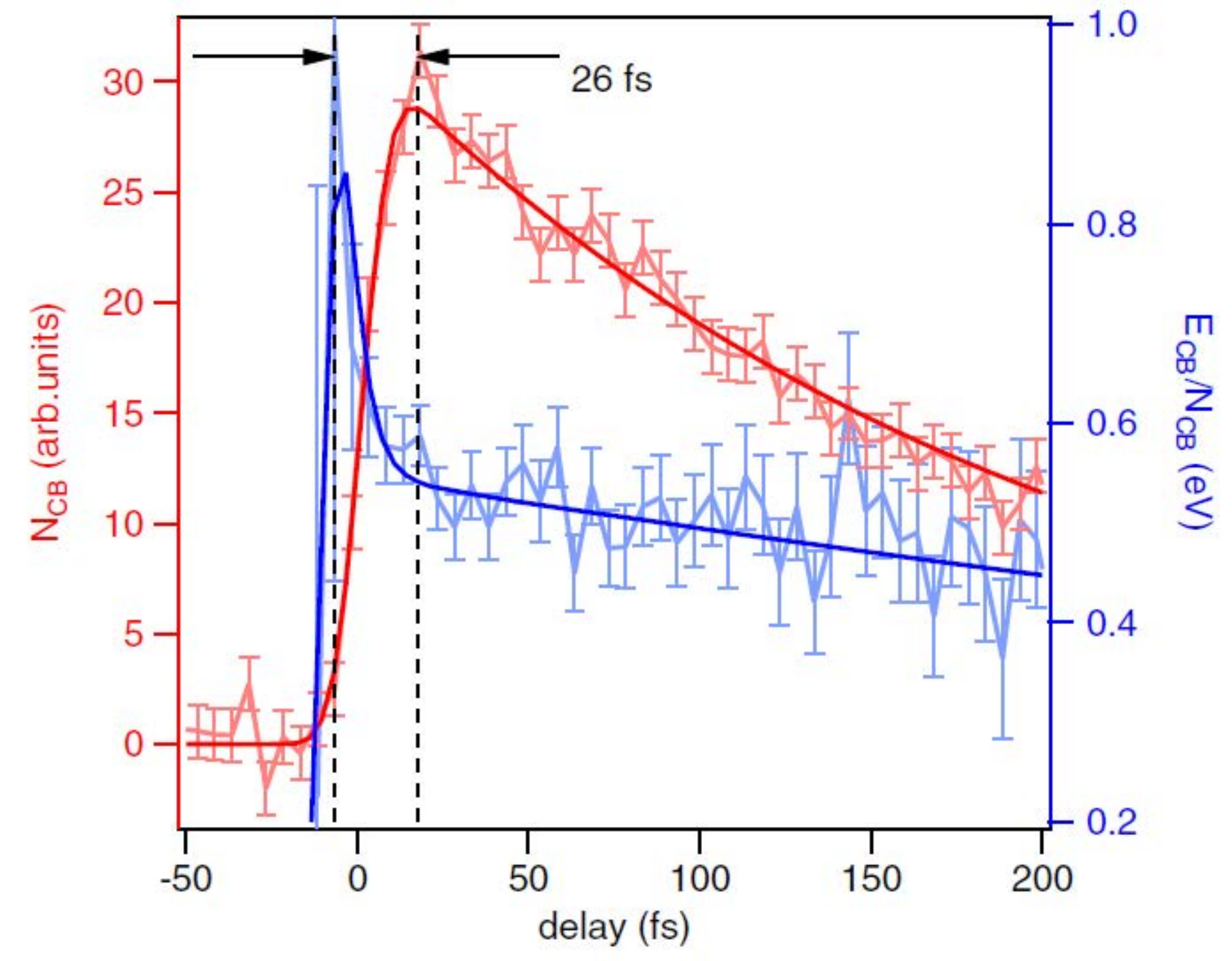}
\caption{Experimental prove for the predominant role of impact ionization in time-and angle-resolved
photoemission (ARPES) measurements.
Comparison between the temporal evolution of the total number
of carriers inside the conduction band (light red) and the
temporal evolution of their average kinetic energy (light blue). Around zero time delay, the energy already decreases
while the number of carriers keeps revealing the importance of impact excitation.
Figure taken from Ref. \cite{gierz15}. }
\label{fig5_arpes}
\end{figure}

Figure \ref{fig4_cm_exp} shows an excellent agreement between theory and experiment with respect to the appearance of the carrier multiplication, its quantitative values as well as its qualitative dependence on the pump fluence \cite{ploetzing14}.  
Since Auger processes are sensitive to the excited carrier density, the appearance of CM strongly varies in different excitation regimes. 
 Depending on the pump fluence and the delay time after the optical pulse, we can clearly distinguish both in experiment and theory two distinct regions characterized by $\rm{CM}>1$ (purple) and $\rm{CM}<1$ (orange). 
We find the largest CM values at low pump fluences in the range of $\unit[0.1]{\mu J cm^{-2}}$.  The CM appears on a timescale of several picoseconds. At intermediate pump fluences up to approximately $\unit[1]{\mu J cm^{-2}}$, we observe a smaller carrier multiplication on a much shorter timescale. Here, IE still prevails over AR, however the asymmetry between these two processes due to Pauli blocking is restricted to a shorter time range, since the number of scattering partners is increased accelerating the relaxation dynamics and leading to a faster equilibration between the IE and AR processes \cite{winzer12b}. In the strong excitation regime with pump fluences larger than $\unit[1]{\mu J cm^{-2}}$, the states are highly occupied and Pauli blocking preferes AR bringing carriers back to the valence band. As a  result, the carrier density decreases and we find a \textit{negative} carrier multiplication $\rm{CM}<1$, cf. Fig. \ref{fig4_cm_exp}.

In addition to the multi-color pump-probe measurements, Gierz and co-workers \cite{gierz15} could demonstrate the presence of a strong impact excitation in a time-and angle-resolved
photoemission spectroscopy study. Extreme-ultraviolet pulses were used to track the number of excited electrons and their kinetic energy. In a time window of approximately 25 fs after absorption of the
pump pulse,  a clear increase of carrier density and a simultaneous decrease of the
average carrier kinetic energy was observed directly (cf. Fig. \ref{fig5_arpes}) revealing that relaxation is dominated by impact excitation.

\section{Carrier multiplication in doped graphene}

Most graphene samples are characterized by a non-zero doping, i.e. the Fermi energy lies above (n doping) or below the Dirac point (p doping). In this section, we investigate the influence of an extrinsic carrier density on the dynamics of optically excited carriers. Since doping breaks the symmetry between electrons and holes, we extend the graphene Bloch equations by deriving separate equations for the time evolution of electron occupations  $\rho^{e,c}_{\bf k}$ in the conduction and and hole occupations  $\rho^{h,v}_{\bf k}$ in the valence band. Furthermore, we include dynamic screening of the Coulomb potential accounting for the increased carrier density.  Our calculations reveal that for n-doped (p-doped) graphene, electron (hole) relaxation is, as expected, faster resulting in a thermalized hot Fermi distribution already after $\unit[30]{fs}$. This can be traced back to the increased number of scattering partners for electrons or holes depending on the type of doping \cite{kadi15}. 

In the presence of doping, we have to reconsider the definition of charge carriers. For n-doping, we can either count electrons above the Dirac point or above the Fermi level (for holes accordingly in p-doped samples). 
 For optical measurements, probing vertical carrier transitions the Dirac point is the more appropriate reference. Then, the carrier density reads
\begin{align}
 \mathcal{N}_{\text{CM}}= \frac{\sigma_s \sigma_v}{L^2} \sum_{\kk, \lambda=e,h} \rho_\kk^{\lambda}, \label{eq:ndef}
\end{align}
where $\sigma_s$ ($\sigma_v$) denotes the spin (valley) degeneracy and $L^2$ the graphene area.  
In contrast to optical measurements, for electric transport phenomena hot electrons around the Fermi level play the crucial role. Thus, charge carriers are defined with  respect to the Fermi level, i.e. for n-doped graphene the upper Dirac cone is split into $\rho_{\kk}^{h,c}=1-\rho_{\kk}^c$ for $k<k_F$ and $\rho_{\kk}^{e,c}=\rho_{\kk}^c$ for $k>k_F$ with the
Fermi momentum $k_F$. The bottom cone remains unaffected with $\rho_{\kk}^{h,v}\equiv \rho_{\kk}^h=1-\rho_\kk^v$. The according \textit{hot} carrier density $\mathcal{N}_{\text{hCM}}$ is given by
\begin{align}
  \mathcal{N}_{\text{hCM}} = \frac{\sigma_s \sigma_v}{L^2} \Big( \sum_\kk \rho_\kk^h+\sum_\kk^{k<k_F} \rho_\kk^{h,c}+\sum_\kk^{k>k_F} \rho_\kk^{e,c}\Big).\label{nhCM}
\end{align}
Note that symmetric results are obtained for n- and p-doped graphene. For undoped graphene, both definitions of 
carrier density and carrier multiplication are equivalent for symmetry reasons.

The definition of the charge carrier density has a direct impact on the appearance of carrier multiplication. Having the Dirac point as the reference,  carrier multiplication occurs due to  Auger scattering bridging the valence and the conduction band. In the following, we label this process as carrier multiplication (CM). Having the Fermi level
as the reference, carrier multiplication occurs via Coulomb-induced intraband scattering bridging
the states below and above the Fermi level, cf. the inset in Fig. \ref{fig7_hcm}(a). According to literature \cite{tielrooij13}, we label this process as hot carrier multiplication (hCM).

\begin{figure}[t!]%
\centering
\includegraphics*[width=\columnwidth]{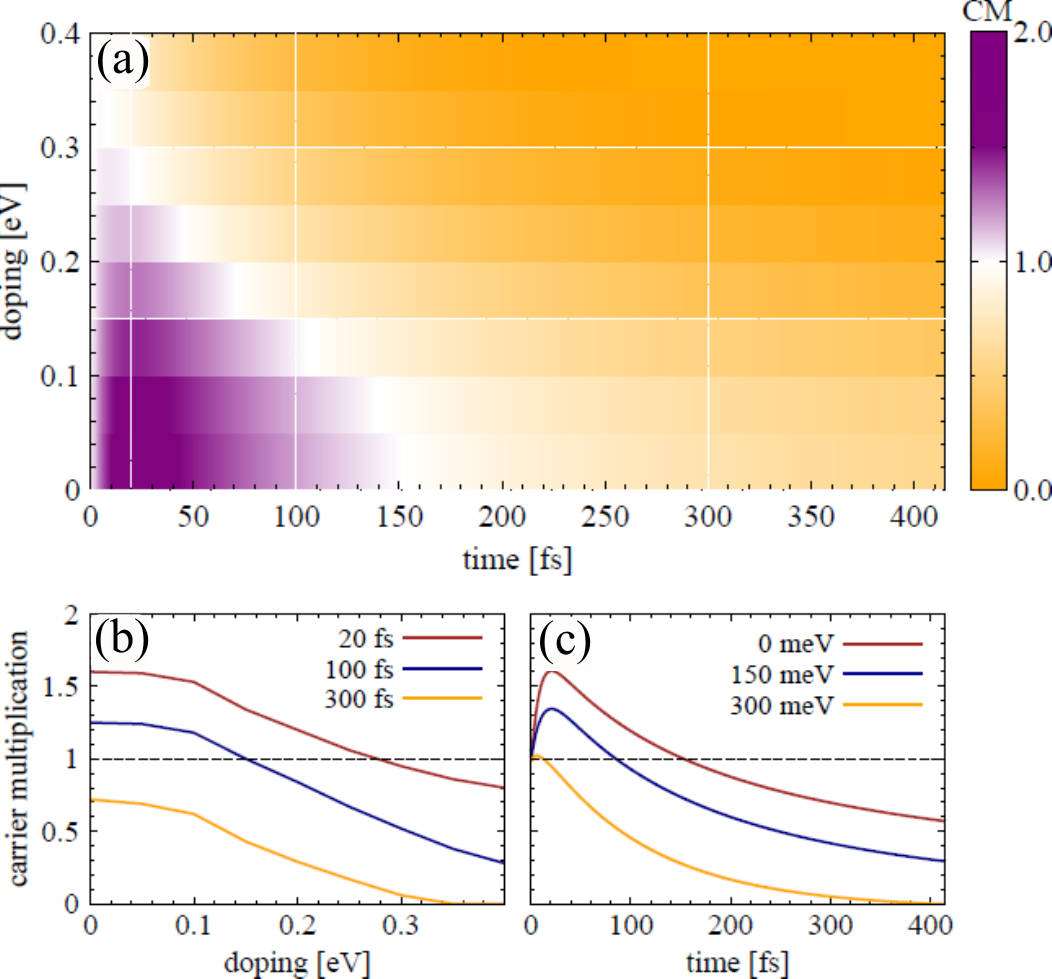}
\caption{Temporal evolution of the doping-dependent CM for a fixed absorbed pump fluence $\varepsilon_{\text{abs}}=\unit[0.3]{\mu J cm^{-2}}$. 
(b) Doping-dependent CM for three fixed time delays and (c) the temporal evolution of CM for different doping levels. 
Figure taken from Ref. \cite{kadi15}. }
\label{fig6_cm}
\end{figure}

We find that  doping reduces the efficiency of CM, cf. Fig. \ref{fig6_cm}(a). Considering an absorbed pump fluence of $\varepsilon_{\text{abs}}=0.3\mu J$ cm$^{-2}$, the highest CM factor of $1.7$ is obtained in the limiting case of undoped graphene, cf. Fig.  \ref{fig6_cm}(b). It appears on a timescale of up to \unit[150]{fs} for undoped graphene and becomes significantly shorter with increasing doping, cf. Fig. \ref{fig6_cm}(c). For doping values higher than \unit[300]{meV}, CM completely vanishes. This behavior can be traced back to the  strongly efficient impact excitation at low doping, which is a result of the large gradient in carrier occupation around the Dirac point. This changes for highly doped graphene presenting optimal conditions for the inverse process of Auger recombination.

\begin{figure}[t!]%
\centering
\includegraphics*[width=\columnwidth]{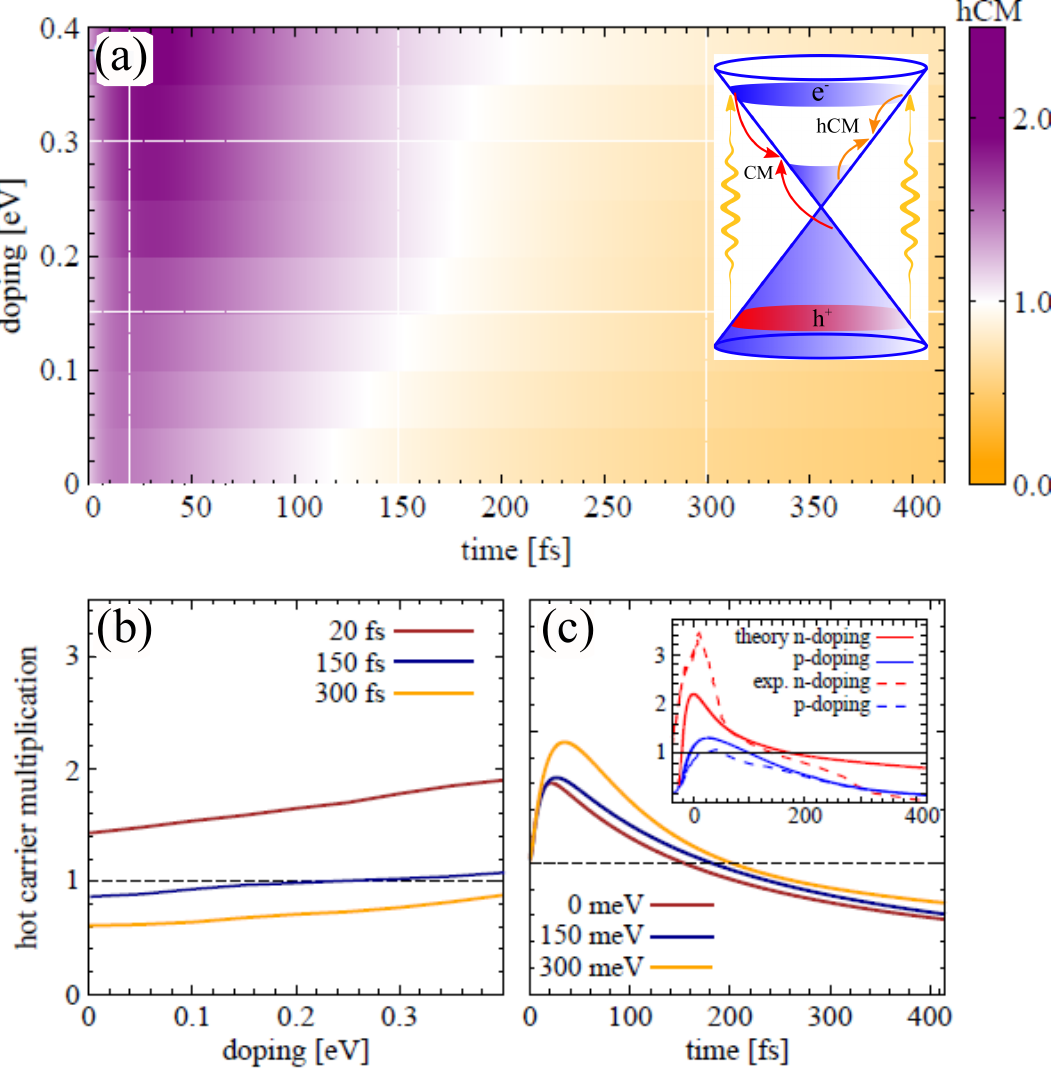}
\caption{The same investigation as in Fig. \ref{fig6_cm} for the case of hot carrier  multiplication. 
The inset in (a) illustrates the processes responsible for CM and  hCM in  n-doped graphene.  
The inset in (f) shows a direct comparison between theoretically predicted (solid lines)  and experimentally measured (dashed lines) data \cite{hofmann15} for: (i) n-doped graphene ($E_F =\unit[380]{meV}$) with an absorbed pump fluence of $\varepsilon_{\text{abs}}=\unit[0.5]{\mu J cm^{-2}}$    and (ii) p-doped graphene ($E_F = \unit[240]{meV}$) with a fluence of
$\varepsilon_{\text{abs}}=\unit[1.5]{\mu J cm^{-2}}$. 
Figure adapted from Ref. \cite{kadi15}. }
\label{fig7_hcm}
\end{figure}

In contrast, we predict a clear increase of the hCM with doping, cf. Fig. \ref{fig7_hcm}(a). We obtain hCM factors of up to approximately $2$ occurring on a timescale of  \unit[200]{fs} for highly doped graphene with $E_F =$ \unit[300]{meV}. 
There is nearly a linear dependence between hCM and doping, cf. Fig. \ref{fig7_hcm}(b). 
The intraband Coulomb scattering around the Fermi energy is responsible for the appearance of hCM. With increasing doping, the spectral region for these \textit{intraband Auger processes} shifts to higher density of states. As a result, they become generally more efficient. The probability for \textit{intraband IE processes} (inset in Fig. \ref{fig7_hcm}(a)) is given by $\rho^{h,c}(1-\rho^{e,c})$, which is initially large compared to the probability for \textit{intraband AR processes} scaling with $\rho^{e,c}(1-\rho^{h,c})$. The initial strong imbalance between IE and AR gives rise to the observed pronounced hCM. 

Applying our microscopic approach, we can explain the recently performed time-resolved ARPES measurements on n- and p-doped graphene \cite{hofmann15}. A direct comparison of theoretically predicted and experimentally obtained values for hot carrier multiplication is shown in the inset of Fig. \ref{fig7_hcm}(c). 
The theory captures well all features observed in the experiment, such as the clearly higher hCM for n-doped graphene. This difference is not due to the type of doping, but can rather be explained by the differences in the applied pump fluence and the doping value of the investigated graphene samples. Hot carrier multiplication increases almost linearly with doping (cf. Fig. \ref{fig7_hcm}(b)) and becomes less efficient at high pump fluences (similarly to the case of CM, cf. Fig. \ref{fig4_cm_exp}). As a result, the n-doped graphene sample shows a much more pronounced hCM, since its Fermi level is significantly higher and since the experiment has been performed at a clearly smaller pump fluence 
compared to the p-doped graphene sample.

\section{Carrier multiplication in Landau-quantized graphene}

Carrier multiplication holds the potential to increase the power conversion
efficiency of photovoltaic devices. However, due to the absence of a bandgap
and competing phonon-induced recombination, the extraction of charge
carriers remains a substantial challenge. A  strategy has been recently suggested to circumvent this drawback and to benefit from the gained charge carriers by introducing a Landau quantization offering a tunable bandgap \cite{wendler14}.
In the presence of magnetic fields the electronic bandstructure of graphene drastically changes. The
Dirac cone collapses into discrete non-equidistant Landau levels, which can be externally tuned by changing
the magnetic field \cite{haldane88,sadowski06,plochocka08,orlita08,miller09}. In contrast to conventional materials, specific Landau levels are selectively addressable using circularly polarized light.

\begin{figure}[t!]%
\centering
\includegraphics*[width=\columnwidth]{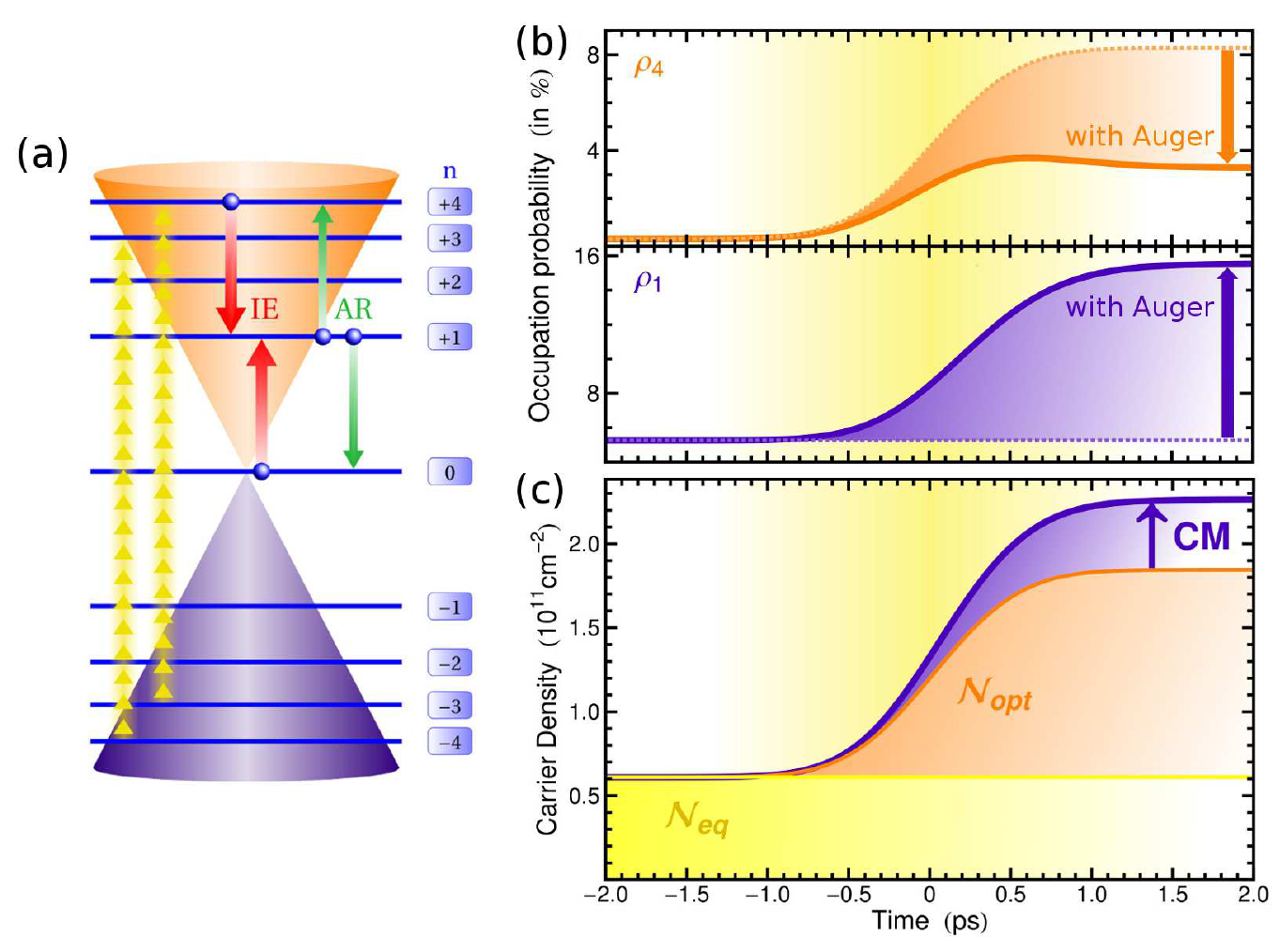}
\caption{(a) Illustration of the low-energetic Landau levels in graphene with the Dirac cone in the background, where the yellow arrows indicate the optical excitation.
Carriers excited to $\text{LL +4}$ can participate in impact excitation (IE), which dominates over the inverse
process of Auger recombination (AR).   (b) Temporal evolution of the
occupations $\rho_{+4}$ and $\rho_{+1}$ during an optical excitation. The dashed lines represent
the dynamics in the absence of Auger scattering. The yellow region
illustrates the width of the pump pulse. (c) Temporal evolution of the total carrier density
 $\mathcal{N}$ (purple line), the optically induced carrier density $\mathcal{N}_{opt}$ (orange line), and the constant equilibrium carrier density $\mathcal{N}_{\rm{eq}}$ (yellow line). Efficient impact excitation leads to a carrier multiplication of approximately $1.3$. Figure taken from Ref. \cite{wendler14}.}
\label{fig8_LL}
\end{figure}

The first experiment measuring the carrier dynamics
in Landau-quantized graphene was performed by Plochocka et al. in
2009 \cite{plochocka09}. Investigating rather high-energetic Landau
levels ($n\sim100$), they found a suppression of Auger processes
as a consequence of the non-equidistant level spacings suggesting
an overall suppression of Auger processes in Landau-quantized graphene
\cite{plochocka09}.
However, due to the $E \propto \sqrt{n}$ dependence, we can always find Landau levels $n$ that are actually equidistant suggesting that Auger scattering might be important in certain situation. This has been confirmed in a  recent polarization-resolved pump-probe experiment addressing the crucial impact of Auger scattering to the low-energetic Landau levels \cite{winnerl15}.

Recently, we have suggested a pumping scheme to open up Auger scattering channels and to achieve a carrier multiplication in Landau-quantizede graphene, cf. Fig. \ref{fig8_LL}(a). The strategy is to excite
charge carriers to $\text{LL +4}$, which induces an energy conserving
scattering process including the transitions $\text{LL +4}\rightarrow\text{LL +1}$
and $\text{LL 0}\rightarrow\text{LL +1}$. This process is denoted
as impact excitation (IE), since it effects the excitation of an additional
charge carrier. Due to Pauli blocking, we expect the inverse process of Auger recombination (AR) to be suppressed.

To investigate the impact of Auger scattering and to prove whether CM appears, we calculate the temporal evolution of the carrier occupation of the involved 
Landau levels $\text{LL +4}$ and $\text{LL +1}$,  cf. Fig. \ref{fig8_LL}(b).
The applied optical pulse is characterized by  a width $\sigma_{\text{FWHM}}^{\text{exp}}=\unit[1]{ps}$ (yellow shaded region),
pump fluence $\varepsilon_{\text{pf}}=\unit[10^{-2}]{\mu Jcm^{-2}}$,
and an energy $ \unit[280]{meV}$ matching the transitions 
$\text{LL} \mp 4 \rightleftarrows\text{LL} \pm 3$
at a magnetic field of $B=\unit[4]{T}$. The system is considered to
be at room temperature and an impurity-induced Landau level broadening
of $\Gamma^{\text{imp}}=\unit[7]{meV}$ is assumed. The dashed lines
in Fig. \ref{fig8_LL}(b) represent the dynamics without Auger
scattering. In this case, the occupation $\rho_{+4}$ simply increases
during the optical excitation, while $\rho_{+1}$ stays constant, since it is not optically excited.
Switching on Auger processes (solid lines) results in a considerable transfer of occupation from
$\text{LL +4}$ to $\text{LL +1}$. The total increase of $\rho_{+1}$ (about $10\%$) exceeds the decrease
of $\rho_{+4}$ (about $5\%$) by a factor of $2$. This reflects the predominant process of impact excitation inducing the transitions $\text{LL +4}\rightarrow\text{LL +1}$ and $\text{LL 0}\rightarrow\text{LL +1}$, cf. the red arrows in Fig. \ref{fig8_LL}(a).

\begin{figure*}[t!]%
\centering
\includegraphics*[width=\textwidth]{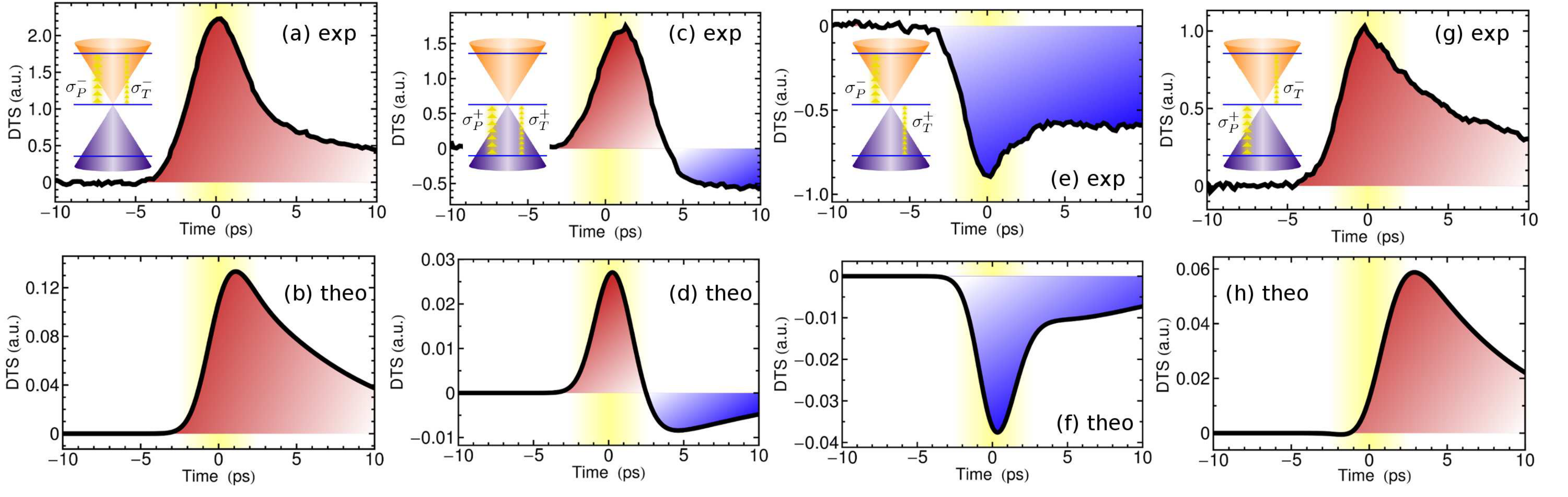}
\caption{Direct comparison of experimentally measured (upper panel) and theoretically predicted (lower panel) polarization-resolved differential transmission for pumping and probing the same transition [(a)-(d)] and different transitions [(e)-(h)]. The corresponding pump ($\sigma_{\text{P}}$) and probe (or test $\sigma_{\text{T}}$) pulses are sketched in the insets. The yellow shaded areas in the background illustrate the width of the pump pulse. 
Figure adapted from Ref. \cite{wendler15}.}
\label{fig9_LL_exp}
\end{figure*}

Figure  \ref{fig8_LL}(c) shows the temporal evolution of the carrier density revealing a clear increase
due to the efficient impact excitation. As a result,  a carrier multiplication  of 1.3 appears in Landau-quantized graphene. Note that the maximal CM that can be achieved within the proposed pumping scheme is 1.5. 
However, an optical excitation into a higher energetic Landau level is conceivable and is expected to induce higher values of the CM. In general, a low pump fluence is
advantageous for CM, since a stronger optical excitation results in
an increased asymmetry between the competing IE and AR scattering
channels, leading to a faster equilibration of the corresponding scattering
rates and reducing the time frame in which IE generates additional
charge carriers. Moreover, changes of the initial occupations result
in an increasing CM with lower temperatures and higher magnetic fields.
Additionally, the CM increases with the magnetic field, as the effective
pumping strength is reduced.
Finally, a higher value of the CM is obtained for larger Landau level
broadenings. 

The theoretically predicted importance of Auger scattering in Landau-quantized graphene is remarkable due to the non-equidistant level separation.  Luckily,  the carrier occupation of single LLs can be directly addressed via polarization-dependent pump-probe experiments allowing an experimental investigation of the importance of Auger channels.
Using circularly polarized light of a specific energy, we can selectively pump and probe transitions
between the energetically lowest Landau levels $\text{LL -1}$, $\text{LL 0}$
and $\text{LL +1}$.
This results in four possibilities to combine pump
and probe pulse polarization, cf. Fig. \ref{fig9_LL_exp}.
Considering only the optical excitation, we expect a positive differential transmission signal (DTS), if the pump $\sigma_P$ and the probe (test) pulse $\sigma_T$ have the same polarization.  Here, the excitation of charge carriers due to the pump pulse should lead to an absorption bleaching of the probe pulse due to the increased Pauli blocking. In contrast, using an opposite polarization
for the pump and the probe pulse, Pauli blocking is expected to be reduced giving rise to  an absorption enhancement and a negative DTS.

The upper and lower panel of Fig. \ref{fig9_LL_exp} illustrate
the experimental and theoretical results, respectively, for the four configurations of pump and probe
pulse polarization applying a pulse with a width of $\sigma_{\text{FWHM}}^{\text{exp}}=\unit[2.7]{ps}$
and an energy of $\unit[75]{meV}$. Comparing the data with
the expectation based on the occupation change induced by the optical
excitation, a qualitative difference is observed both in experiment and theory in the case of pumping
with $\sigma^{-}$-polarized radiation (Figs. \ref{fig9_LL_exp}(e)-(h)): While the configuration $\sigma_{\text{P}}^{-},\,\sigma_{\text{T}}^{-}$ shows an initial increase
(as expected) which is followed by an unexpected sign change to the negative region,
the measured behavior in the configuration $\sigma_{\text{P}}^{-},\,\sigma_{\text{T}}^{+}$
is completely contrary to what we would expect. 

To explain these surprising results, we need to take into account Auger scattering.
Moreover, to achieve an agreement between experiment and theory, a
finite doping needs to be introduced to break the electron-hole symmetry.
In an undoped system, the two configurations $\sigma_{\text{P}}^{+},\,\sigma_{\text{T}}^{+}$
and $\sigma_{\text{P}}^{-},\,\sigma_{\text{T}}^{-}$ (and likewise
$\sigma_{\text{P}}^{+},\,\sigma_{\text{T}}^{-}$ and $\sigma_{\text{P}}^{-},\,\sigma_{\text{T}}^{+}$)
would yield the same result. The assumption of a finite doping
is further supported by experimental studies showing that multilayer
epitaxial graphene samples grown on the C-terminated face of SiC (such
a sample is used in the experiment) have a finite n-doping due to
a charge transfer from the SiC substrate \cite{sprinkle09,sun10b}.
Therefore, for the theoretical calculations a Fermi energy of $\varepsilon_{\text{F}}=\unit[28]{meV}$ has been assumed.  
\begin{figure}[b!]%
\centering
\includegraphics*[width=\columnwidth]{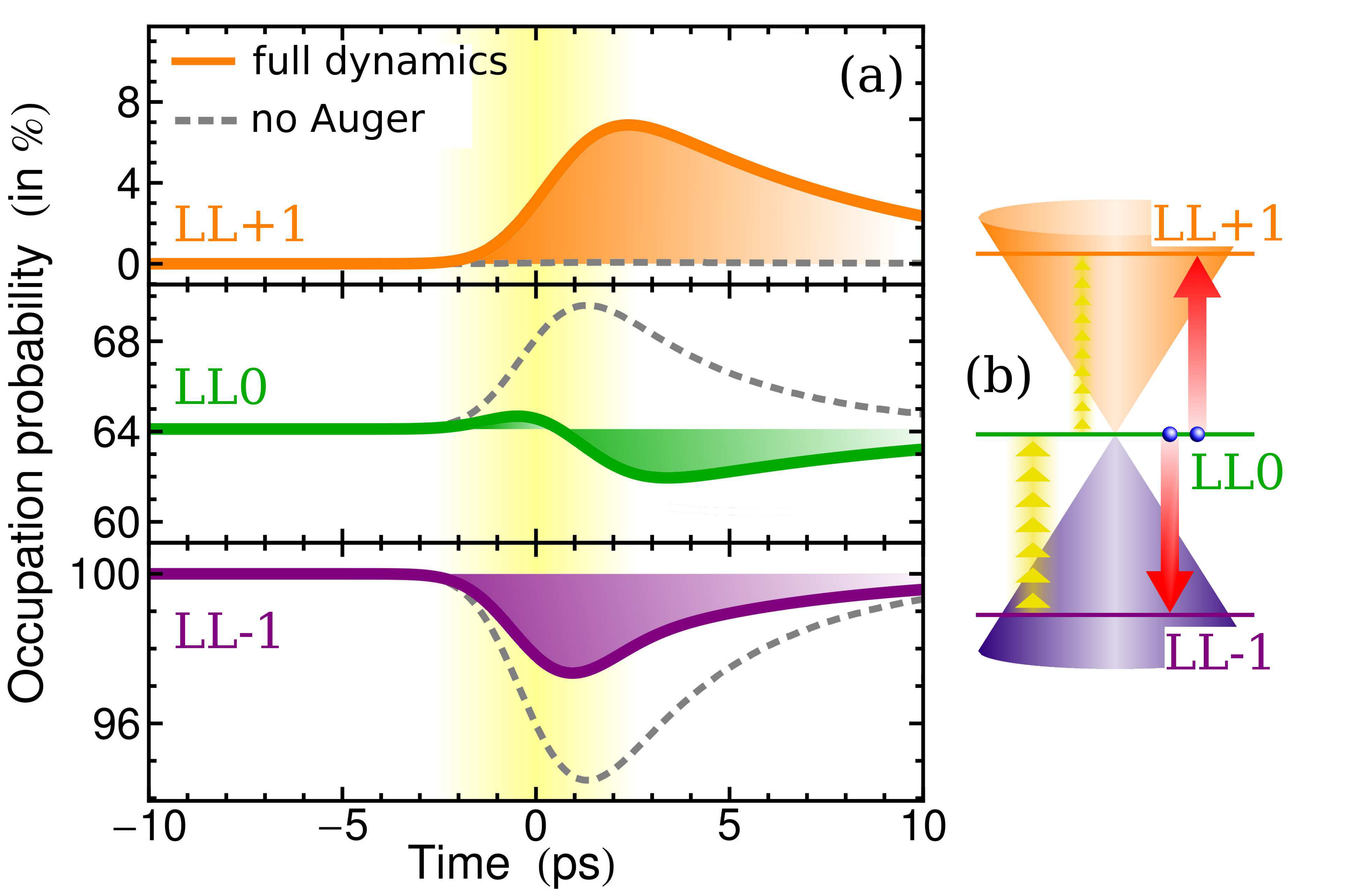}
\caption{ (a) Temporal evolution of the three energetically lowest Landau involved in the Auger processes from Fig. \ref{fig9_LL_exp}
after a $\sigma^{-}$-polarized excitation. (b) The sketch illustrates the interplay of Auger scattering and optical excitation explaining that Auger scattering can even lead to a depopulation of the zeroth Landau level even though carriers are pumped into this level. Figure adapted from Ref. \cite{winnerl15}.}
\label{fig10_LL_exp_sol}
\end{figure}

To understand the unexpected DTS sign in Fig. \ref{fig9_LL_exp}(g), we investigate the temporal evolutions of the
involved Landau level occupations $\rho_{+1}$, $\rho_{0}$ and $\rho_{-1}$
after the application of a $\sigma^{-}$-polarized pump pulse, cf. Fig. \ref{fig10_LL_exp_sol}(a). To identify the role of  Auger scattering, we compare the full dynamics with the calculation without Auger processes, cf. the dashed gray line. Note that  carrier-phonon scattering has been phenomenologically included to match the experimentally observed fast decay rates at long times.  Although the transition $\text{LL -1}\rightarrow\text{LL 0}$
is optically induced, the occupation $\rho_{0}$ shows only a minor increase in the beginning,
then it starts even to decrease already before the center of the pulse
is reached. This means that although we optically pump carriers into the zeroth Landau level, its population decreases. This surprising result can be explained by extremely efficient Auger scattering, which  induces the
transitions $\text{LL 0}\rightarrow\text{LL -1}$ and $\text{LL 0}\rightarrow\text{LL +1}$,
thereby resulting in a quick depopulation of $\rho_{0}$, cf. Fig. \ref{fig10_LL_exp_sol}(b). The crucial role of the Auger scattering becomes apparent by comparing the temporal evolution of $\rho_{0}$ to the case without Coulomb scattering
(cf. dashed gray lines in Fig. \ref{fig10_LL_exp_sol}(a). Here, as expected, the population of the zeroth Landau level increases during the entire time of the optical excitation.

As a result, the polarization-dependent pump-probe experiments provide
proof for efficient Auger scattering in Landau-quantized graphene. The unexpected DTS sign  emerges since Auger scattering depopulates
the zeroth Landau level faster than it is filled by optical excitation.
This surprising effect appears, when the pumping efficiency is decreased
due to an enhanced Pauli blocking as a result of a finite doping.

In summary, we have reviewed the current research on carrier multiplication in undoped, doped, and Landau-quantized graphene. We find that Auger scattering bridging the valence and the conduction band dominates the carrier dynamics in graphene. In particular, impact excitation is the predominant relaxation channel under certain conditions resulting in a significant carrier multiplication. 

We thank our experimental collaborators, in particular Stephan Winnerl and Manfred Helm (Helmholtz-Zentrum Dresden-Rossendorf) as well Daniel Neumeier and Heinrich Kurz (AMO Aachen). Furthermore, we acknowledge financial support from the Deutsche Forschungsgemeinschaft (DFG) through the priority programm 1458, the EU Graphene Flagship (contract no. CNECT-ICT-604391), and the Swedish Research Council (VR).


\begin{thebibliography}{10}

\bibitem{nozik02}
A.J Nozik.
\newblock {Quantum dot solar cells}.
\newblock {\em Physica E}, 14:115 -- 120, 2002.

\bibitem{malic13}
Ermin Malic, Stephan Winnerl, and Andreas Knorr.
\newblock {\em Graphene and Carbon Nanotubes: Ultrafast Optics and Relaxation
  Dynamics}.
\newblock Wiley-{VCH}, 1 edition, May 2013.

\bibitem{winzer10}
Torben Winzer, Andreas Knorr, and Ermin Malic.
\newblock Carrier multiplication in graphene.
\newblock {\em Nano Letters}, 10(12):4839--4843, December 2010.

\bibitem{winzer12b}
Torben Winzer and Ermin Mali{\'c}.
\newblock Impact of auger processes on carrier dynamics in graphene.
\newblock {\em Physical Review B}, 85(24):241404, June 2012.

\bibitem{schaller04}
R.~D. Schaller and V.~I. Klimov.
\newblock {High Efficiency Carrier Multiplication in PbSe Nanocrystals:
  Implications for Solar Energy Conversion}.
\newblock {\em Phys. Rev. Lett.}, 92:186601, 2004.

\bibitem{schaller05}
R.~D. Schaller, V.~M. Agranovich, and V.~I. Klimov.
\newblock {High-efficiency carrier multiplication through direct
  photogeneration of multi-excitons via virtual single-exciton states}.
\newblock {\em Nature Phys.}, 1:189--194, 2005.

\bibitem{ellingson05}
Randy~J. Ellingson, Matthew~C. Beard, Justin~C. Johnson, Pingrong Yu, Olga~I.
  Micic, Arthur~J. Nozik, Andrew Shabaev, and Alexander~L. Efros.
\newblock {Highly Efficient Multiple Exciton Generation in Colloidal PbSe and
  PbS Quantum Dots}.
\newblock {\em Nano Lett.}, 5:865--871, 2005.

\bibitem{gur05}
Ilan Gur, Neil~A. Fromer, Michael~L. Geier, and A.~Paul Alivisatos.
\newblock {Air-Stable All-Inorganic Nanocrystal Solar Cells Processed from
  Solution}.
\newblock {\em Science}, 310(5747):462--465, 2005.

\bibitem{scholes06}
Gregory~D. Scholes and Garry Rumbles.
\newblock {Excitons in nanoscale systems}.
\newblock {\em Nat. Mater.}, 5(9):683--696, September 2006.

\bibitem{schaller06}
R.~D. Schaller, M.~Sykora, J.~M. Pietryga, and V.~I. Klimov.
\newblock {Seven Excitons at a Cost of One: Redefining the Limits for
  Conversion Efficiency of Photons into Charge Carriers}.
\newblock {\em Nano Lett.}, 6:424--429, 2006.

\bibitem{nair11}
G.~Nair, L.-Y. Chang, S.~M. Geyer, and M.~G. Bawendi.
\newblock {Perspective on the Prospects of a Carrier Multiplication Nanocrystal
  Solar Cell}.
\newblock {\em Nano Lett.}, 11:2145--2151, 2011.

\bibitem{gabor09}
Nathaniel~M. Gabor, Zhaohui Zhong, Ken Bosnick, Jiwoong Park, and Paul~L.
  {McEuen}.
\newblock Extremely efficient multiple electron-hole pair generation in carbon
  nanotube photodiodes.
\newblock {\em Science}, 325(5946):1367--1371, September 2009.

\bibitem{baer10}
Roi Baer and Eran Rabani.
\newblock Can impact excitation explain efficient carrier multiplication in
  carbon nanotube photodiodes?
\newblock {\em Nano Letters}, 10(9):3277--3282, September 2010.

\bibitem{wang10}
Haining Wang, Jared~H. Strait, Paul~A. George, Shriram Shivaraman, Virgil~B.
  Shields, Mvs Chandrashekhar, Jeonghyun Hwang, Farhan Rana, Michael~G.
  Spencer, Carlos~S. Ruiz-Vargas, and Jiwoong Park.
\newblock Ultrafast relaxation dynamics of hot optical phonons in graphene.
\newblock {\em Applied Physics Letters}, 96(8):081917--081917--3, February
  2010.

\bibitem{gabor13}
N.~M. Gabor.
\newblock Impact excitation and electronâ€“hole multiplication in graphene and
  carbon nanotubes.
\newblock {\em Acc. Chem. Res.}, 46:1348, 2013.

\bibitem{kanemitsu13}
Y.~Kanemitsu.
\newblock {Multiple Exciton Generation and Recombination in Carbon Nanotubes
  and Nanocrystals}.
\newblock {\em Acc. Chem. Res.}, 46:1358--1366, 2013.

\bibitem{brida13}
D.~Brida, A.~Tomadin, C.~Manzoni, Y.~J. Kim, A.~Lombardo, S.~Milana, R.~R.
  Nair, K.~S. Novoselov, A.~C. Ferrari, G.~Cerullo, and M.~Polini.
\newblock {Ultrafast collinear scattering and carrier multiplication in
  graphene}.
\newblock {\em Nature Commun.}, 4:1987, 2013.

\bibitem{wendler14}
Florian Wendler, Andreas Knorr, and Ermin Malic.
\newblock {Carrier multiplication in graphene under Landau quantization}.
\newblock {\em Nature Commun.}, 5:3703, 2014.

\bibitem{ploetzing14}
T.~Pl\"otzing, T.~Winzer, E.~Malic, D.~Neumaier, A.~Knorr, and H.~Kurz.
\newblock {Experimental Verification of Carrier Multiplication in Graphene}.
\newblock {\em Nano Lett.}, 14(9):5371--5375, 2014.
\newblock PMID: 25144320.

\bibitem{landsberg93}
P.~T. Landsberg, H.~Nussbaumer, and G.~Willeke.
\newblock {Band-band impact ionization and solar cell efficiency}.
\newblock {\em J. Appl. Phys.}, 74:1451--1452, 1993.

\bibitem{shockley61}
William Shockley and Hans~J. Queisser.
\newblock Detailed balance limit of efficiency of p-n junction solar cells.
\newblock {\em Journal of Applied Physics}, 32(3):510--519, March 1961.

\bibitem{basko13}
D.~M. Basko.
\newblock Effect of anisotropic band curvature on carrier multiplication in
  graphene.
\newblock {\em Phys. Rev. B}, 87:165437, 2013.

\bibitem{kadi15}
F.~Kadi, T.~Winzer, A.~Knorr, and E.~Malic.
\newblock Impact of doping on the carrier dynamics in graphene.
\newblock {\em Sci. Rep.}, 5:16841, 2015.

\bibitem{gierz15}
I.~Gierz, F.~Calegari, S.~Aeschlimann, M.~Chavez Cervantes, C.~Cacho, R.~T.
  Chapman, E.~Springate, S.~Link, U.~Starke, C.~R. Ast, and A.~Cavalleri.
\newblock Tracking primary thermalization events in graphene with photoemission
  at extreme timescales.
\newblock {\em Phys. Rev. Lett.}, 115:086803, 2015.

\bibitem{hofmann15}
Jens~Christian Johannsen, S{\o}ren Ulstrup, Alberto Crepaldi, Federico Cilento,
  Michele Zacchigna, Jill~A. Miwa, Cephise Cacho, Richard~T. Chapman, Emma
  Springate, Felix Fromm, Christian Raidel, Thomas Seyller, Phil D.~C. King,
  Fulvio Parmigiani, Marco Grioni, and Philip Hofmann.
\newblock Tunable carrier multiplication and cooling in graphene.
\newblock {\em Nano Letters}, 15(1):326--331, 2015.
\newblock PMID: 25458168.

\bibitem{winnerl15}
M.~Mittendorff, F.~Wendler, E.~Malic, A.~Knorr, M.~Orlita, M.~Potemski,
  C.~Berger, W.~A. de~Heer, H.~Schneider, M.~Helm, and S.~Winnerl.
\newblock {Carrier dynamics in Landau-quantized graphene featuring strong Auger
  scattering}.
\newblock {\em Nature Phys.}, 11:75--81, 2015.

\bibitem{knorr96}
A.~Knorr, S.~Hughes, T.~Stroucken, and S.~W. Koch.
\newblock Theory of ultrafast spatio-temporal dynamics in semiconductor
  heterostructures.
\newblock {\em Chemical Physics}, 210:27 -- 47, 1996.

\bibitem{malic11b}
Ermin Malic, Torben Winzer, Evgeny Bobkin, and Andreas Knorr.
\newblock Microscopic theory of absorption and ultrafast many-particle kinetics
  in graphene.
\newblock {\em Physical Review B}, 84(20):205406, November 2011.

\bibitem{malic16}
Ermin Malic, Torben Winzer, Florian Wendler, and Andreas Knorr.
\newblock {\em Microscopic view on ultrafast carrier dynamics in graphene, in
  Optical Properties of Graphene, ed. by R. Binder}.
\newblock World Scientific, 2016.

\bibitem{wendler15}
Florian Wendler, Andreas Knorr, and Ermin Malic.
\newblock Ultrafast carrier dynamics in landau-quantized graphene.
\newblock {\em Nanophotonics}, 4(1):224--249, 2015.

\bibitem{torben_phd}
T.~Winzer.
\newblock {\em Ultrafast Carrier Relaxation Dynamics in Graphene}.
\newblock PhD thesis, Technische Universit\"at Berlin, July 2013.

\bibitem{winzer14}
Torben Winzer and Ermin Malic.
\newblock Carrier multiplication and optical gain in graphene.
\newblock Proc. SPIE, page 8984, 2014.

\bibitem{tielrooij13}
K.~J. Tielrooij, J.~C.~W. Song, S.~A. Jensen, A.~Centeno, A.~Pesquera,
  A.~Zurutuza~Elorza, M.~Bonn, L.~S. Levitov, and F.~H.~L. Koppens.
\newblock Photoexcitation cascade and multiple hot-carrier generation in
  graphene.
\newblock {\em Nature Physics}, 9:248, February 2013.

\bibitem{haldane88}
F.~D.~M. Haldane.
\newblock {Model for a Quantum Hall Effect without Landau Levels:
  Condensed-Matter Realization of the "Parity Anomaly"}.
\newblock {\em Phys. Rev. Lett.}, 61:2015--2018, 1988.

\bibitem{sadowski06}
M.~L. Sadowski, G.~Martinez, M.~Potemski, C.~Berger, and W.~A. de~Heer.
\newblock {Landau Level Spectroscopy of Ultrathin Graphite Layers}.
\newblock {\em Phys. Rev. Lett.}, 97:266405, 2006.

\bibitem{plochocka08}
P.~Plochocka, C.~Faugeras, M.~Orlita, M.~L. Sadowski, G.~Martinez, M.~Potemski,
  M.~O. Goerbig, J.-N. Fuchs, C.~Berger, and W.~A. de~Heer.
\newblock {High-Energy Limit of Massless Dirac Fermions in Multilayer Graphene
  using Magneto-Optical Transmission Spectroscopy}.
\newblock {\em Phys. Rev. Lett.}, 100:087401, 2008.

\bibitem{orlita08}
M.~Orlita, C.~Faugeras, P.~Plochocka, P.~Neugebauer, G.~Martinez, D.~K. Maude,
  A.-L. Barra, M.~Sprinkle, C.~Berger, W.~A. de~Heer, and M.~Potemski.
\newblock {Approaching the Dirac Point in High-Mobility Multilayer Epitaxial
  Graphene}.
\newblock {\em Phys. Rev. Lett.}, 101:267601, 2008.

\bibitem{miller09}
David~L. Miller, Kevin~D. Kubista, Gregory~M. Rutter, Ming Ruan, Walt~A.
  de~Heer, Phillip~N. First, and Joseph~A. Stroscio.
\newblock {Observing the Quantization of Zero Mass Carriers in Graphene}.
\newblock {\em Science}, 324:924--927, 2009.

\bibitem{plochocka09}
P.~Plochocka, P.~Kossacki, A.~Golnik, T.~Kazimierczuk, C.~Berger, W.~A.
  de~Heer, and M.~Potemski.
\newblock {Slowing hot-carrier relaxation in graphene using a magnetic field}.
\newblock {\em Phys. Rev. B}, 80:245415, 2009.

\bibitem{sprinkle09}
M.~Sprinkle, D.~Siegel, Y.~Hu, J.~Hicks, A.~Tejeda, A.~Taleb-Ibrahimi,
  P.~Le~F\`evre, F.~Bertran, S.~Vizzini, H.~Enriquez, S.~Chiang,
  P.~Soukiassian, C.~Berger, W.~A. de~Heer, A.~Lanzara, and E.~H. Conrad.
\newblock {First Direct Observation of a Nearly Ideal Graphene Band Structure}.
\newblock {\em Phys. Rev. Lett.}, 103:226803, 2009.

\bibitem{sun10b}
Dong Sun, Charles Divin, Julien Rioux, John~E. Sipe, Claire Berger, Walt~A.
  de~Heer, Phillip~N. First, and Theodore~B. Norris.
\newblock Coherent control of ballistic photocurrents in multilayer epitaxial
  graphene using quantum interference.
\newblock {\em Nano Letters}, 10(4):1293--1296, April 2010.

\end{thebibliography}

\end{document}